\documentstyle[epsfig]{mn}
\begin{document}


\title[Observations of HDF with {\em ISO} - III]{Observations of the
{\em Hubble Deep Field} with the {\em Infrared Space Observatory} - III:
Source Counts and $P(D)$ Analysis}
\author[Oliver, Goldschmidt, Franceschini {\it et al.}]
{ S.J. Oliver$^1$, 
 P. Goldschmidt$^1$, 
 A. Franceschini$^4$,
 S.B.G. Serjeant$^1$,
\vspace*{0.2cm}\\{\LARGE  
A.N. Efstathiou$^1$,
A. Verma$^1$,
C. Gruppioni$^1$,
N.Eaton$^1$,
R.G. Mann$^1$,
}\vspace*{0.25cm}\\ {\LARGE 
B. Mobasher$^1$,
C.P. Pearson$^1$,
M. Rowan-Robinson$^1$,
T.J. Sumner$^1$, 
}\vspace*{0.2cm}\\{\LARGE 
L. Danese$^2$, 
D. Elbaz$^3$, 
E. Egami$^5$, 
M. Kontizas$^6$, 
A. Lawrence$^7$, 
}\vspace*{0.2cm}\\{\LARGE 
R. McMahon$^8$, 
H.U. Norgaard-Nielsen$^9$, 
I. Perez-Fournon$^{10}$, 
I. Gonzalez-Serrano$^{11}$
}\\
$^1$Astrophysics Group, Imperial College London, Blackett Laboratory,
Prince Consort Road, London SW7 2BZ;\\ 
$^2$SISSA, Via Beirut 2-4, Trieste, Italy;\\
$^3$Service d'Astrophysique, Saclay, 91191, Gif-sur-Yvette, Cedex,
France;\\ 
$^4$Osservatorio Astronomico de Padova, Vicolo dell'Osservatorio 5,
I-35 122, Padova, Italy;\\
$^5$Max-Planck-Institut f\"ur Extraterrestrische Physik,
Giessenbachstrasse, D-8046, Garching bei Munchen, Germany;\\
$^6$Astronomical Institute, National Observatory of Athens, P.O.Box
200048, GR-118 10, Athens, Greece;\\
$^7$Institute for Astronomy, University of Edinburgh, Blackford Hill,
Edinburgh, EH9 3HJ;\\
$^8$Institute of Astronomy, The Observatories, Madingley Road,
Cambridge, CB3 0HA;\\
$^9$Danish Space Research Institute, Gl. Lundtoftevej 7, DK-2800
Lyngby, Copenhagen, Denmark;\\
$^{10}$Instituto Astronomico de Canarias, Via Lactea, E-38200 La
Laguna, Tenerife, Canary Islands, Spain;\\
$^{11}$Instituto de Fisica de Cantabria, Santander, Spain\\
}
\date{Accepted 1997 May 9;
      Received 1997 March 24;
      in original form 1996 Decmber 5}
 
\pagerange{\pageref{firstpage}--\pageref{lastpage}}
\pubyear{1997}
\volume{847}

\label{firstpage}

\maketitle
\begin{abstract}

We present source counts at 6.7 $\mu$m and 15 $\mu$m from our maps of
the Hubble Deep Field region, reaching 38.6$\mu$Jy at 6.7 $\mu$m and
255$\mu$Jy at 15 $\mu$m. These are the first ever extra-galactic number
counts to be presented at $6.7\mu$m and are 3 decades fainter than
{\em IRAS} at 12 $\mu$m.  Both source counts and a $P(D)$ analysis suggest we
have reached the {\em ISO} confusion limit at 15 $\mu$m: this will have
important implications for future space missions.  These data provide
an excellent reference point for other ongoing {\em ISO} surveys.  A
no-evolution model at 15 $\mu$m is ruled out at $>3\sigma$, while two
models which fit the steep {\em IRAS} 60 $\mu$m counts are acceptable. This
provides important confirmation of the strong evolution seen in {\em IRAS}
surveys.  One of these models can then be ruled out from the 6.7 $\mu$m
data.

\end{abstract}
\begin{keywords}
galaxies:$\>$formation - 
infrared: galaxies - surveys - galaxies: evolution - 
galaxies: star-burst -
galaxies: Seyfert

\end{keywords}


\section{Introduction} 

{\em IRAS} galaxy surveys at 60 $\mu$m  have consistently
provided good evidence for a population of star forming galaxies
evolving with a strength comparable to AGN.  This has been confirmed
by numerous studies of count distributions and redshift surveys from
0.6Jy to 50mJy (Hacking and Houck 1987; Saunders {\em et al.} 1990;
Lonsdale {\em et al.} 1990; Oliver {\em et al.} 1995; Gregorich
{\em et al.} 1995; Bertin, Dennefeld and Moshir {\em et al.} 1997).

This evolving population discovered by {\em IRAS} could have very important
implications for cosmological studies.  In particular these objects
are likely to contribute strongly to the star formation history of the
Universe.  [Other incidental issues include the possibly significant
impact such objects could have on the cosmological far infrared
background e.g. Oliver {\em et al.} (1992) and Franceschini {\em et al.}
(1991).]  The populations seen by
{\em IRAS} are mostly relatively low redshift ($z<0.2$);
deeper {\em ISO} surveys such as this provide a longer baseline in redshift
giving a better handle on the nature of the
evolution.

This paper will discuss the source counts from our maps of the
Hubble Deep Field (HDF, Williams {\em et al.} 1996).  Our observations
have been described by Serjeant {\em et al.} (1997; 
Paper I) and the
source extraction described by Goldschmidt {\em et al.} (1997; 
Paper II): 
a total of 27 sources were found at 6.7 $\mu$m, and 22 at 15
 $\mu$m. Further papers discuss the associations with optical
galaxies (Mann {\em et al.}  1997; 
Paper IV) and the models for
spectral energy distributions and the star formation history
Rowan-Robinson {\em et al.} (1997; 
Paper V).

\section{Observed Source Counts at 6.7 and 15 $\mu$m}

In Paper II we detected 7 sources in our 6.7 $\mu$m maps and 19 in our
15 $\mu$m maps using a well defined source detection algorithm.  To
convert these source lists into source counts requires an estimate of
the area within which a source of observed flux ($S_{\nu}$) could have
been detected.

To estimate this
we need to know the minimum flux ($S_{\rm lim}$) detectable at any position.
The source detection algorithm requires $m$ pixels to have  intensity
$I(x,y)>T(x,y)$, where $T(x,y)$ is the threshold intensity.  The flux
of resulting detections is estimated using an estimated local
background intensity $B(x,y)$.  Assuming a well determined point
spread function (PSF),
$P(x-x_0,y-y_0)$, this algorithm gives us a detection limit
\begin{equation}
S_{\rm lim}(x_0,y_0)={\rm max}_m \left(\frac{T(x,y)-B(x,y)}{P(x-x_0,y-y_0)}\right) 
\end{equation}
where ${\rm max}_m$ is a function giving the  $m$th largest value.
$T$ is defined for various areas in Paper II
and $B$ is determined by running the sky annulus across 
the full image.

The PSF is the only uncertainty in this estimate of the survey areas,
we thus decided to investigate this in some detail.  One estimate for
$P$ comes from the standard ESA products.  The PSF is estimated at
$3\times3$ sub pixel positions for each filter and lens position.  We
drizzled these images together to produce a single PSF for each of
our two observing modes. For both bands, the ESA PSF contains a large
amount of flux in the wings.  These wings may be caused by scattered
light or other data reduction features in the ESA PSF.
In any case we cannot use
the PSF for analysing the effective area since only 20 per cent
of the total flux is within the Airy disk, so we calculate a `revised'
ESA PSF which is background subtracted in the same way as were the
sources and normalised to 1 within the source aperture.  
Since
our observations involved long integrations, in which jitter
might significantly blur the PSF, and also because our data reduction
did not take into account field distortions,
we decided to estimate $P$ empirically from the HDF observations
themselves.
 $P$ was estimated by summing the intensities from a
number of sources in an square aperture 7 pixels to a side (i.e. 7
arcsec at 6.7 $\mu$m and 21 arcsec at 15 $\mu$m).  The relatively
small aperture at 6.7 $\mu$m was necessary to avoid including more
than one source.
At 15
 $\mu$m we excluded objects near the boundaries and the fainter sources
leaving a sample of 6 sources, while at
6.7 $\mu$m we used all sources in the complete sample.   
We then normalised such that $\sum P=1$ over the
aperture.  Some parameters of the PSFs discussed are summarised in 
Table \ref{psf}

\begin{table}
\caption{Characteristics of various Point Spread Functions: 
column 1, peak intensity;
column 2, full width half maximum; column3 flux within twice the
nominal Airy disk
diameter; column 4, intensity in $m$th brightest pixel (pixels are
1 arcsec at 6.7 $\mu$m and 3 arcsec at 15 $\mu$m}\label{psf}

\begin{tabular}{lllll}\hline

PSF                    & $P_{\rm max}$ & FWHM & $S_{2D}$ & max$_m(P)$\\
                       & /arcsec$^{-2}$  & /arcsec &       &
                       /arcsec$^{-2}$ \\
ESA 6.7 $\mu$m          & 0.038         & 3.0  & 0.68     &  0.023\\
Revised ESA 6.7 $\mu$m  & 0.057         & 3.0  & 1.00     &  0.034\\
Empirical 6.7 $\mu$m    & 0.051         & 4.2  & 1.00     &  0.035\\
ESA 15 $\mu$m           & 3.0e-3        & 5.4  & 0.20     &  7.0e-4\\
Revised ESA 15 $\mu$m   & 1.7e-2        & 5.4  & 1.00     &  3.8e-3\\
Empirical 15 $\mu$m     & 6.2e-3        &10.0  & 1.00     &  3.4e-3\\
\hline
\end{tabular}   

\end{table}

\begin{figure*}
\epsfig{file=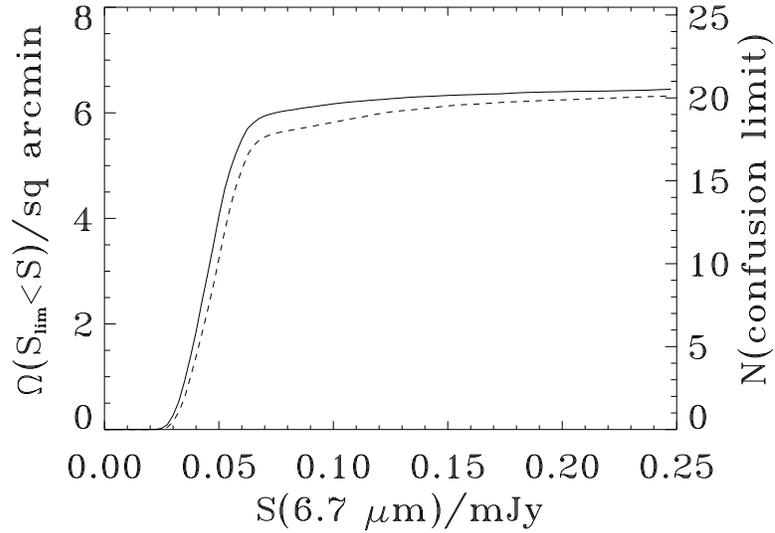, angle=90, width=10cm}
\caption{Area in which the 6.7 $\mu$m map is sensitive 
to sources at a given flux. The dashed line is calculated using the
re-normalised ESA PSF. On the right is the number
of sources that would cause one to exceed the classical confusion
limit of 40 beams per source.}\label{area7}
\end{figure*}

\begin{figure*}

\epsfig{file=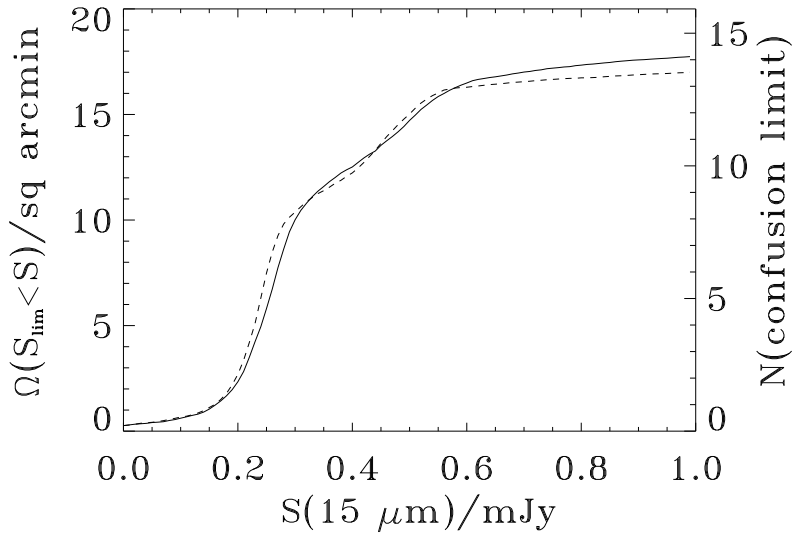, angle=90, width=10cm}
\caption{Area in which the 15 $\mu$m map is sensitive to sources at
a given flux. The dashed line is calculated using the re-normalised
ESA PSF.  On the right is the number of sources that would cause one
to exceed the classical confusion limit of 40 beams per
source.}\label{area15}

\end{figure*}

Figures \ref{area7} and \ref{area15} show the effective
areas of the surveys to sources of a given flux [$\Omega(S_{\rm lim}<S)$].
Notice that that the curves do not pass through the origin
in Figure \ref{area15},  this is because the sources were detected
using a fixed global threshold but the background is estimated
locally, hence a source could in principal be detected in a high 
background region but then assigned zero or even negative flux.

The faintest flux limit that can provide useful information is defined
by the smallest useable area. We choose the smallest useful area to be
200 beams, using Figures \ref{area7} and \ref{area15} this translates
to 38.6 and 255 $\mu$Jy over 1.6 and 6.3 arcmin$^2$ respectively.
These flux limits include 6 of the 7 6.7 $\mu$m sources and 17 of the
19 15 $\mu$m sources.  Allowing a smaller area of 40 beams would have
suggested a flux limit of 30.4 and 161 $\mu$Jy in which case the
faintest 15 $\mu$m source would be excluded and the faintest 6.7 $\mu$m
source would be at the flux limit.

At this point we can examine whether our HDF data approach the
confusion limit of {\em ISO}.  The maximum area available at 15 $\mu$m is
around 18 arcmin$^2$ thus the complete sample from Paper II of 19
sources is below the classical confusion limit 
(more than 1 source every 40 beams)
see Figure \ref{area15}.  As most of the complete $6.7\mu$m sample of
7 objects have fluxes are above 40$\mu$Jy, Figure \ref{area7} shows
that they are confined to an area $> 2$ arcmin$^2$ and thus are above
the classical confusion limit.  Including the supplementary list from
Paper II
provides a combined sample of 27 sources which could well
represent a confused sample as the maximum survey area available is
$\sim 6$ arcmin$^2$.

We use the  areas determined with the empirical PSF to construct the 
observed differential source counts or integral counts
\begin{equation}
dN_i=\sum_{j: dS}\frac{1}{\Omega(S_{\rm lim}<S_j)},
\end{equation}
\begin{equation}
N(>S)=\sum_{j: >S}\frac{1}{\Omega(S_j)}.
\end{equation}

\begin{figure*}
\epsfig{file=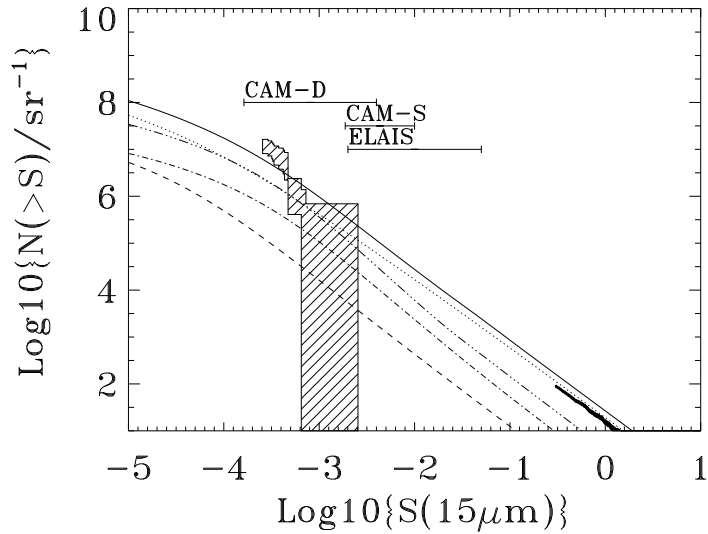, angle=90, width=10cm}
\caption{15 $\mu$m source counts from HDF (hatched region)
with {\em IRAS} 12 $\mu$m counts (thick line) at bright end, 
from this paper and Rush {\em et al.} 1993 
({\em IRAS} data shifted to 15 $\mu$m using cirrus spectrum).  Models 
based on Pearson  and Rowan-Robinson (1996): 
all components solid;
normal galaxies dotted;
evolving star-bursts dash-dot-dot-dot;
Seyferts 1 and 2   dash and dot-dash.
The depths probed by other, forthcoming {\em ISO} surveys are indicated 
(ELAIS: Oliver {\em et al.} 1997;
CAM-D \& CAM-S: e.g. Elbaz 1997; 
Taniguchi: Taniguchi {\em et al.} 1997)
}\label{cpp_15}
\end{figure*}

The integral counts are shown in
Figures \ref{cpp_7} through \ref{af_15} together with models and (in
Figures \ref{cpp_15} and \ref{af_15}) {\em IRAS} counts, which are discussed
later.  The integral counts are shown rather than differential counts
as these are more useful in practice,  however the data points are not
independent and so Poisson $1\sigma$ limits are
shown as a hatched area. Rigorous statistical
comparisons between data and models are best made by assessing the number of 
objects expected at given flux limits and for this we apply the area
as a function of flux to the model; these results are
discussed in Section \ref{comp}

With the exception of the analysis of simulated data sets we
exclude objects in our complete list that are not associated with
optical counterparts in Paper IV.
  This criterion removes one source
at 6.7 $\mu$m and 5 at 15 $\mu$m.

\section{Stellar Source Counts}

Before moving on to discuss the galaxy counts a brief word needs to be
said about the stellar counts.  At 6.7 $\mu$m there are no obvious
stellar candidates in the complete sample.  This is no real surprise
since the HDF was selected to exclude bright stars.  Extrapolations
from the models of Franceschini {\em et al.} (1991) predict 0.38 
star arcmin$^{-2}$ for $S_{6.7}>40\mu$ Jy at these latitudes, 
i.e.  1.9 stars if the HDF area was
not biased against stars, reasonably consistent with our finding none.
At 15 $\mu$m we see one stellar image in the flanking field ({\em ISO}HDF3
J123709.8+621239).  The Franceschini {\em et al.} (1991) stellar model
would predict 0.1 star arcmin$^{-2}$ i.e. around one to two stars in this field
which is quite consistent with our single detection. 
This stellar object has been  exclude from the galaxy counts.

\section{Comparison with Far Infrared Source Count Models}

\subsection{{\em IRAS} Galaxy Counts}

We can use the 12 $\mu$m {\em IRAS} galaxy counts to estimate the bright
counts at 15 $\mu$m using the
Rush {\em et al.} (1993) sample.  In order to estimate the area of
this sample we had to construct a mask to take account of the {\em IRAS}
missing strip and regions of high source density that may have been
excluded or underrepresented.  To this end we applied the QMW {\em IRAS}
Galaxy Catalogue mask
(Rowan-Robinson {\em et al.} 1991); this cut excluded only 43
galaxies, reducing the sample to 850 objects and provided an estimate
of the area of 6.76 sr.    These counts are shown in Figures \ref
{cpp_15} and \ref{af_15}.

\subsection{Model Galaxy Populations}\label{models}

\begin{figure*}
\epsfig{file=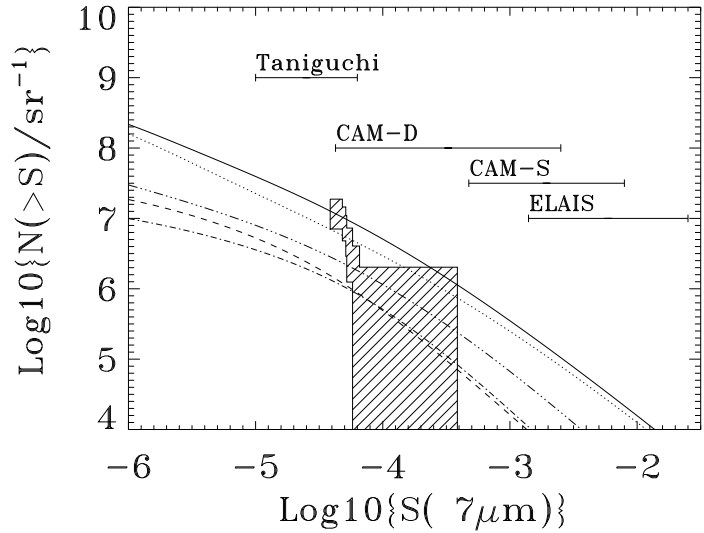, angle=90, width=10cm}
\caption{6.7 $\mu$m source counts from HDF (hatched region).
  Models at 6.7 $\mu$m based on  Pearson  and Rowan-Robinson (1996): 
all components solid;
normal galaxies dotted;
evolving star-bursts dash-dot-dot-dot;
Seyferts 1 and 2 dash and dot-dash.
The depths probed by other, forthcoming {\em ISO} surveys are indicated 
(ELAIS: Oliver {\em et al.} 1997;
CAM-D \& CAM-S: e.g. Elbaz 1997; 
Taniguchi: Taniguchi {\em et al.} 1997)
}\label{cpp_7}
\end{figure*}

We consider two models of galaxy number counts.  Both models have
well defined dust emission spectra specifically to
accurately predict mid to far infrared galaxy distributions.  In
addition both models include strongly evolving components sufficient
to explain the steep number counts at 60  $\mu$m.  The specific
populations and spectral energy distributions in the two models are 
however significantly
different.  Pearson and Rowan-Robinson (1996; PRR) have described a galaxy
population model involving five populations: normal galaxies;
star-burst galaxies; hyper-luminous galaxies; Seyfert 1 galaxies and
Seyfert 2 galaxies.  We have predicted the counts at 6.7 $\mu$m,
12 $\mu$m and 15 $\mu$m using these models ignoring the hyper-luminous
population which will have negligible contribution.  Here star-bursts
and normal galaxies have 60 $\mu$m luminosity functions taken from
Saunders {\em et al.}  (1990), the Seyferts 12  $\mu$m luminosity
functions come from Rush {\em et al.} (1993).  Both Seyferts and star
bursts evolve as $L(z)=L(0)(1+z)^{3.1}$. The SEDs used for these
galaxies are based mainly on {\em IRAS} data and are described in Pearson
and Rowan-Robinson (1996).  The model assumes an $\Omega=1$ cosmology.
These models were shown to provide a good fit to the {\em IRAS} 60 $\mu$m
counts.  As they stand these models would be unable to account for
optical or K-band counts but would require a low $\Omega$ or more
strongly evolving star-burst population.  The integral counts
predicted by this model are shown in Figures \ref{cpp_7} and
\ref{cpp_15}.

\begin{figure*}
\epsfig{file=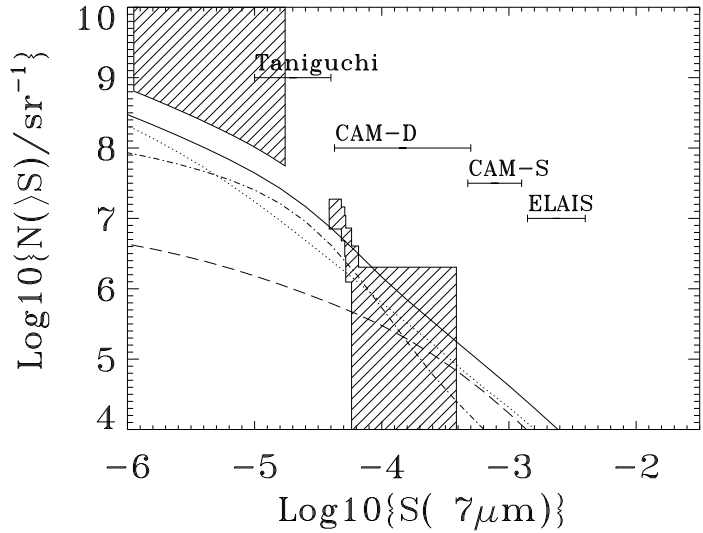, angle=90, width=10cm}
\caption{6.7 $\mu$m source counts from HDF (central hatched regions).  Models 
based on Franceschini {\em et al.} (1994): 
all components solid;
spiral galaxies dotted;
E+s0 dot-dash;
AGN-dash.
Faint end upper limits (hatched) come from the $P(D)$ analysis.
The depths probed by other, forthcoming {\em ISO} surveys are indicated 
(ELAIS: Oliver {\em et al.} 1997;
CAM-D \& CAM-S: e.g. Elbaz 1997; 
Taniguchi: Taniguchi {\em et al.} 1997)
}\label{af_7}
\end{figure*}

\begin{figure*}
\epsfig{file=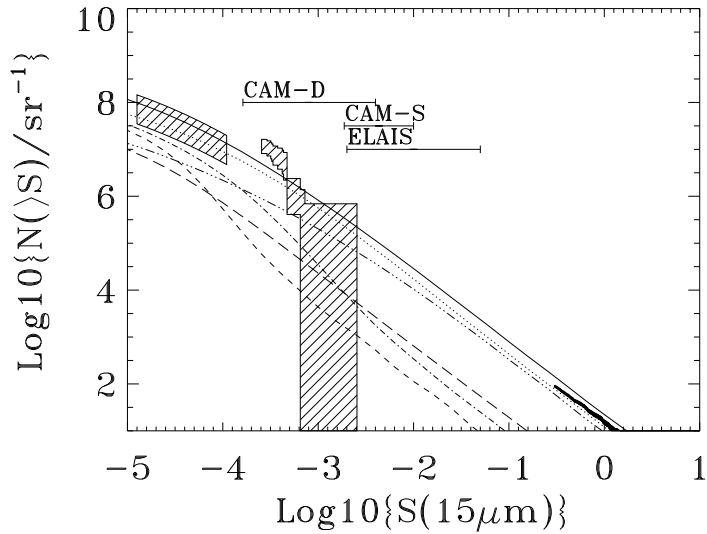, angle=90, width=10cm}
\caption{15 $\mu$m source counts from HDF (central hatched regions)
with {\em IRAS} 12 $\mu$m counts (thick line) at bright end, 
from this paper and Rush {\em et al.} 1993
({\em IRAS} data shifted to 15 $\mu$m using cirrus spectrum).  Models 
based on Franceschini {\em et al.} (1994): 
all components solid;
spiral galaxies dotted;
star-bursts dash-dot-dot-dot;
s0 dot-dash;
AGN-dash
Elliptical galaxies-short dashes.
Faint end constraints (hatched) come from the $P(D)$ analysis.
The depths probed by other, forthcoming {\em ISO} surveys are indicated 
(ELAIS: Oliver {\em et al.} 1997;
CAM-D \& CAM-S: e.g. Elbaz 1997)
}\label{af_15}
\end{figure*}

A second model comes from Franceschini {\em et al.} (1994; AF). The total
counts are modelled as the sum of 5 populations: AGN; star-burst
galaxies; spiral/irregular galaxies; S0 galaxies; and elliptical galaxies.  The
late-type systems (Spirals, Irr and star-bursts) evolve as:
$L(z)=L(0)e^{2 \tau(z)}$ (where $\tau$ is the look-back time), 
in an open universe ($q_0=0.15$).  The
early-type systems (Ellipticals, S0) evolving according to
Franceschini {\em et al.} (1994), i.e. assuming that a bright phase of
active star-formation at $z\sim 2-4$ is obscured by dust quickly
produced by the first stellar generations. This same model accounts in
some detail for the sub-mm background as estimated by Puget {\em et
al.} (1996).  The AGN number density is set by the local {\em IRAS} samples
at 12  $\mu$m (see Rush et al. 1993). The evolution is calibrated so as
to produce the hard X-ray background with a suitable distribution of
the dust/gas absorbing column densities.  The integral
counts predicted by this model are shown in Figures \ref{af_7} and
\ref{af_15}.

The first impression from these Figures is
of a surprising agreement between the models and the data,
considering that in the case of the 15 $\mu$m data the predictions were
made from data 3 decades brighter, while  at 6.7 $\mu$m there
were no previous data with which to normalise the models.
In Section \ref{comp} we will compare the models and observed counts
more rigorously, but first will discuss possible
biases in this data.

\subsection{Simulations of the Observed Source Counts}
\label{bias}

In any comparison of source counts with models, it is vital to discuss
possible sources of bias that might be present.

The first issue is the reliability of the detections.  Spurious
sources arise particularly through non-Gaussian noise features
remaining after the data reduction.  Using optical associations
Paper IV estimates lower limits to the reliability for the complete samples
of 71 and 68 per cent at 6.7 $\mu$m and 15 $\mu$m.  Thus the reliability
could be comparable to the Poisson errors and could be important to
this analysis. For the direct comparison with the models we have
excluded the non-associated detections, thus, by this estimate of
reliability our list is 100 per cent reliable.  This method may
introduce incompleteness, if we have thrown out genuine sources where
the true association has an erroneously low likelihood (as might
happen if our positional errors are not well understood).  An
alternative approach is to include all detections and make a full
assessment of the reliability.  This requires a good estimate of the
noise distribution.

Confusion between faint objects of high source density relative to the
beam is an important factor in these surveys.  Co-addition of
overlapping source profiles means observed fluxes may be an overestimate
of {\em true} single-source fluxes.  On the other hand faint sources
will tend to be merged together so decreasing their numbers. We can
only estimate this bias with knowledge of the source counts at fainter
fluxes.  It is thus more appropriate to incorporate confusion to the models
than to try and correct the data.

Flux errors will cause objects to be scattered from fainter fluxes
above the flux limit and vice versa.  Since there are more fainter
sources these are preferentially scattered over the flux limit into
our survey.  This `Eddington' bias (Eddington 1913), also depends
sensitively on counts at faint magnitudes and so will be included in
the models rather than in the data.

The total integral counts predicted at both wavelengths from the first
of these models have been used to create 100 synthetic galaxy images
using the empirical PSF.  These source images were added to the
background maps created with random pixel positions (see Paper I). 
These background maps include uncorrelated noise and sky
backgrounds but have real sources (or structure) suppressed to a
negligible level. Some sources of correlated noise may also be
suppressed in these background maps and this is a slight limitation to
the simulations.

These resulting images have been passed through the same source
detection algorithm as was applied in Paper II
thereby simulating all
the possible biases discussed above.  The only remaining possible
source of incompleteness is from uncertainties in our PSF which may
warrant further investigation.  

The average number of sources detected from these simulations are
listed in Tables \ref{cppexp7} and \ref{cppexp15}.  They do not appear
to be significantly different from the model counts on which they are
based.  Although the simulations are only made for the first model
this has similar differential counts to the second at faint fluxes so
we would expect similar behaviour.  In neither band do we see any
appreciable discrepancies between the models and the simulations over
the flux ranges including our data and conclude that any biases are
small or cancel each other out.

\subsection{Comparison of Models}\label{comp}

\begin{table}
\caption{Expected number of sources in {\em ISO} HDF Survey above
38.6$\mu$Jy at 6.7 $\mu$m, taking into account the
areal sensitivity in Figure \protect\ref{area7}
using the PRR model.
The number of associated sources observed above 
this flux limit was 5.  For the simulation we did not exclude
the one non-associated source above our flux limit 
since the simulation accounts
for spurious sources.
The second column indicates the
number of sources expected at fluxes brighter than our brightest 
source.  The final column is the probability of the observed
fluxes being drawn from the model distribution, estimated
using a K-S test}\label{cppexp7}

\begin{tabular}{lllll}\hline
Component & $N$ & $N$  & K-S \\
    & $(S>38.6\mu\rm{Jy})$ &$(S>65.7\mu\rm{Jy})$ & /\% \\

         Normal:& 3.13 & 2.27 &  0.44\\
     Star-bursts:& 1.27 & 0.92 &  0.48\\
      Seyfert 1:& 0.66 & 0.44 &  1.30\\
      Seyfert 2:& 0.60 & 0.42 &  0.59\\
          Total:& 5.67 & 3.95 &  0.53\\
    Simulation:& 4.37 & 2.86 &  0.73\\
\hline
\end{tabular}

\end{table}

\begin{table}
\caption{Expected number of sources in {\em ISO} HDF Survey above
38.6$\mu$Jy at 6.7 $\mu$m, taking into account the
areal sensitivity in Figure \protect\ref{area7}
using the AF model.
The number of sources observed above 
this flux limit was 5.  The second column indicates the
number of sources expected at fluxes brighter than our brightest 
source.  The final column is the probability of the observed
fluxes being drawn from the model distribution, estimated
using a K-S test}\label{afexp7}

\begin{tabular}{lllll}\hline
Component & $N$ & $N$  & K-S \\
    & $(S>38.6\mu\rm{Jy})$ &$(S>65.7\mu\rm{Jy})$ & /\% \\
          Spiral:& 1.33&  0.78 &  3.56\\
      Elliptical/S0:& 2.10&  0.98 & 15.78\\
             AGN:& 0.34&  0.26 &  0.21\\
           Total:& 3.77&  2.03 &  6.87\\
\hline
\end{tabular}

\end{table}

\begin{table}
\caption{Expected number of sources in {\em ISO} HDF Survey above
255$\mu$Jy at 15 $\mu$m, taking into account the
areal sensitivity in Figure \protect\ref{area15}
using the PRR.
The number of associated sources observed above 
this flux limit was 11.  For the simulation we did not exclude
the five non-associated source above our flux limit 
since the simulation accounts
for spurious sources. The second column indicates the
number of sources expected at fluxes brighter than our brightest 
source.  The final column is the probability of the observed
fluxes being drawn from the model distribution, estimated
using a K-S test}\label{cppexp15}

\begin{tabular}{lllll}\hline
Component & $N$ & $N$  & K-S \\
    & $(S>255\mu\rm{Jy})$ &$(S>726\mu\rm{Jy})$ & /\% \\

         Normal & 2.71 & 0.94 &  4.22\\
    Star-bursts & 2.81 & 0.89 &  5.83\\
      Seyfert 1 & 0.73 & 0.22 &  9.00\\
      Seyfert 2 & 0.78 & 0.26 &  4.57\\
          Total & 7.03 & 2.32 &  5.26\\
     Simulation & 8.37 & 1.88 &  2.31\\
\hline
\end{tabular}

\end{table}

\begin{table}
\caption{Expected number of sources in {\em ISO} HDF Survey above 255$\mu$Jy
at 15 $\mu$m, taking into account the areal sensitivity in Figure
\protect\ref{area15} using the AF model.  The number of
sources observed above this flux limit was 11.  The second column
indicates the number of sources expected at fluxes brighter than our
brightest source.  The final column is the probability of the observed
fluxes being drawn from the model distribution, estimated using a K-S
test}\label{afexp15}

\begin{tabular}{lllll}\hline
Component & $N$ & $N$  & K-S \\
    & $(S>255\mu\rm{Jy})$ &$(S>726\mu\rm{Jy})$ & /\% \\

          Spiral:& 4.55 & 1.59 &  3.57\\
       Star-burst:& 1.26 & 0.52 &  1.25\\
      Elliptical:& 0.03 & 0.01 & 47.21\\
              S0:& 0.77 & 0.13 & 56.45\\
             AGN:& 0.25 & 0.07 & 14.93\\
           Total:& 6.87 & 2.33 &  4.76\\
\hline
\end{tabular}

\end{table}

Using the area of the survey to a given flux limit (Figures
\ref{area7} and \ref{area15}) and the number count models of Section
\ref{models} we can estimate the number of sources of any given type
that we would expect in this survey, these numbers are summarised in
Tables \ref {cppexp7},\ref{cppexp15},\ref{afexp7},\ref{afexp15}.  Both
models are consistent with the total number of associated galaxies.
The simulation at 15 $\mu$m predicts that there should be 8.37 objects
but our non-associated list includes 16 galaxies.  This is a
marginally significant excess and may indicate that either our
association requirements are too harsh or that our simulated noise is
not realistic; improved understanding of the {\em ISO} data in the near
future will clarify this.  Interestingly the PRR model predicts a
significant fraction of objects ($\sim 70$ per cent) of the objects
should be at fluxes brighter than our brightest source.  This
discrepancy is also seen in the integral count plots (Figure
~\ref{cpp_7}), where we can also see a apparent difference in slope
between the model and the data.

To test whether discrepancies in the count slope are significant we
perform a K-S test to assess the probability of the observed fluxes being
drawn from a count distribution with the same shape as the models (or
single components of the models), these probabilities are given in
Tables\ref {cppexp7},\ref{cppexp15},\ref{afexp7},\ref{afexp15}.  These
results suggest that the PRR model at 6.7
 $\mu$m is ruled out with more than 99 per cent significance (using
either the simulations or raw models), this is mainly due to the expected
fraction of bright objects.  This model cannot be ruled out with more
than 95 per cent confidence at the other wavelength although the
simulation implies a more significant discrepancy, this may again
suggest problems with our simulations or association criteria.

In this test the AF model can also not be rejected with any
high level of confidence (more than around 95 percent) at either wavelength.

It thus appears that these data are insufficient to rule out either
model at 15 $\mu$m.  The PRR model can be ruled
out at the shorter wavelength due to the high fraction of bright
galaxies predicted.  It may be that revisions to the K-corrections of
the ``normal'' component using improved SEDs may be sufficient to
improve this model.  Possibly problems with the ``normal'' component
SED are also suggested by the noticeable over-prediction of the {\em IRAS} 15
 $\mu$m counts in the model.

A ``no evolution'' version of the PRR star-burst
component would predict only 0.28  galaxies 15 $\mu$m i.e. 4.5 in
total.  Since we comfortably  detect 11 galaxies such a model can be ruled out
at the 3 $\sigma$ level on integral counts alone.

\section{$P(D)$ Analysis}

Since our maps are close to the {\em ISO} confusion limit it is sensible to
examine the low level fluctuations in the maps on a statistical level.
This allows us to investigate source count models below the flux level
at which individual sources can be resolved.  To this end we explore the
distribution of deviations in flux in a given aperture from the mean
level, the $P(D)$ distribution.  This analysis avoids the need for any
source detection algorithm.  The fluctuations in a map about the mean
intensity consist of two components.  The first is the noise which will
have both positive and negative values, and will have some distribution function
which we shall assume is Gaussian.  The second is true fluctuations.
These might arise from sources, cosmological background or Galactic
foreground.  In all cases this contribution will always be positive
and thus skew a symmetric noise distribution.  In this case we assume
that any major asymmetric component arises from extra-galactic
sources.

Firstly we select sky regions and construct a histogram of flux
deviations in square apertures.  These histograms are fitted with both
Gaussian distributions functions and the expected distribution
functions from the source count models.  These model distribution
functions are calculated following Franceschini {\em et al.}  (1989)
and using the AF model for the source
counts discussed above (the differential counts of this model are
similar to those from the PRR model below
the source detection limit so the AF model
results will be similar for both,  and the 
model provides a better fit to the count distribution at 6.7 $\mu$m).
The differential count
distribution is first convolved with a model PSF (in this case a
Gaussian PSF is assumed) to give the response function to single
sources in the selected aperture.  The $P(D)$ can then be calculated
assuming the sources are distributed on the sky as a Poisson process.

Figure \ref{pd15} shows the deflection distribution ($D$ in $\mu$Jy)
obtained from a 72x72 matrix of pixels derived from the inner portion
of the drizzled mosaic of Paper I.
We estimate a sky standard
deviation in the inner map {\em outside} obvious sources of
$\sigma=16-17 \mu$Jy/aperture.  (dotted line in the Figure
\ref{pd15}).  The aperture is 6x6 arcsec$^2$ area, enclosing
a disc with diameter equal to the FWHM of the theoretical PSF.  
The mean intensity was 0.4 mJy arcsec$^{-2}$.

The continuous thick line is the convolution of the Gaussian noise
with a model $P(D)$ based on the counts appearing in Figure
\ref{af_15}.  The convolved curve provides a very good fit to the data
(reduced $\chi^2 \sim 1$). A simple Gaussian cannot fit the data, even
if the width is increased to $\sigma=18-19 \mu$Jy aperture$^{-1}$ with reduced
$\chi^2>1.5$.  Thus there is a clear positive signal in the
$P(D)$. This means that the background in an extremely deep {\em ISO}
exposure at $15\mu$m is structured. Such structure is entirely
consistent with being due to an smooth extrapolation of the {\em ISO}
source counts observed above the flux threshold, and confirms the
counts. 

This $P(D)$ analysis allows us to further constrain the models.  The
shape of the AF model counts was fixed and the
normalisation allowed to vary.  These constraints are illustrated in
Figure \ref{af_15}

A similar analysis at 6.7 $\mu$m demonstrated that a simple Gaussian
noise model was sufficient to provide a good fit to the observed
$P(D)$.  This allowed only an upper limit on the normalisation of the
counts to be determined.

\begin{figure*}
\vspace{4cm}
\caption{$P(D)$ analysis at 15 $\mu$m (see text)}\label{pd15}
\end{figure*}

\section{Discussion of Populations}\label{pops}

So far we have not made any use of the fact that our survey was
conducted in the Hubble Deep Field and Flanking Fields where there is
exceptional photometric and spectroscopy information available.  These
data allows us to determine the populations detected by {\em ISO}. 
It is still instructive to compare observed and 
predicted populations as this may provide clues for improvements to
the models.

One of the 6.7 $\mu$m sources ({\em ISO}HDF2 J123646.4+621406) selected at 6.7
 $\mu$m has broad emission lines (Paper IV). This is compatible with the
predicted numbers of AGN in both of the models we have considered.
Of the remaining 4 associated objects only one is compatible with a
normal cirrus spectrum galaxy and has an elliptical morphology, the
others are better fitted with star-burst spectra and have spiral
morphologies (Paper V).

\begin{figure*}
\epsfig{file=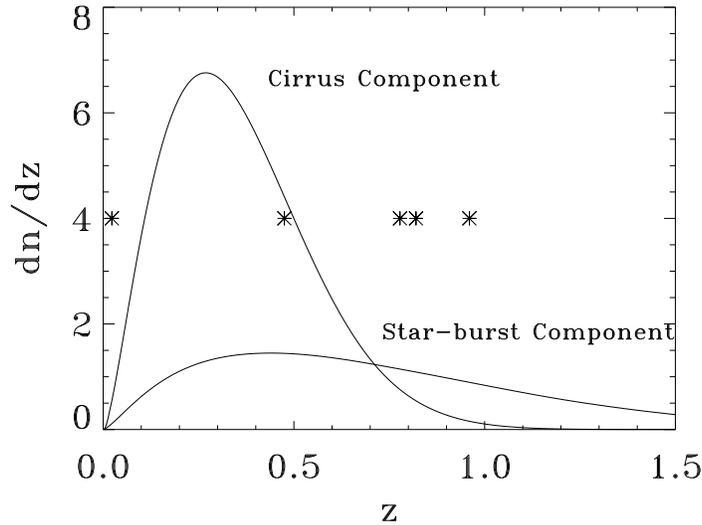, angle=90, width=10cm}
\caption{Expected redshift distribution at 6.7 $\mu$m within the {\em ISO} HDF areas
(accounting for the area dependent flux limit) for the star-burst
and cirrus component of the PRR model.
Over-plotted at arbitrary $y$ position are the redshifts (photometric or
spectroscopic) for
the 5 objects with reliable associations in Paper IV.  The
object at a redshift of 0.96 has broad lines in the optical spectrum.
The probability that these data are drawn from the expected
distribution estimated using the K-S test is 0.54 or 018 if the
broad-lined object is included.}\label{cppz7}
\end{figure*}

Excluding AGN, the first of the models (PRR) suggests that at 6.7
 $\mu$m our sample should have a majority (71 per cent) of normal
spiral galaxies with around 28 per cent star-burst galaxies.  The
expected probability of getting the observed one or fewer cirrus
galaxies in a sample of 4 is $\sim7$ per cent.  Although the
statistics are very poor this casts further doubt on the validity of
PRR model at this wavelength.  The classification of SEDs in Paper V
is based mainly on the optical/IR luminosities.  Extra information
comes from the expected redshift distribution of these normal galaxies
and star-bursts which is shown in Figure \ref{cppz7}.  A K-S test
cannot rule out the possibility that the observed redshifts are drawn
from the model distribution even if the redshift of the broad lined
object is included: the probability of the data being drawn from the
model distribution is 0.18.

\begin{figure*}
\epsfig{file=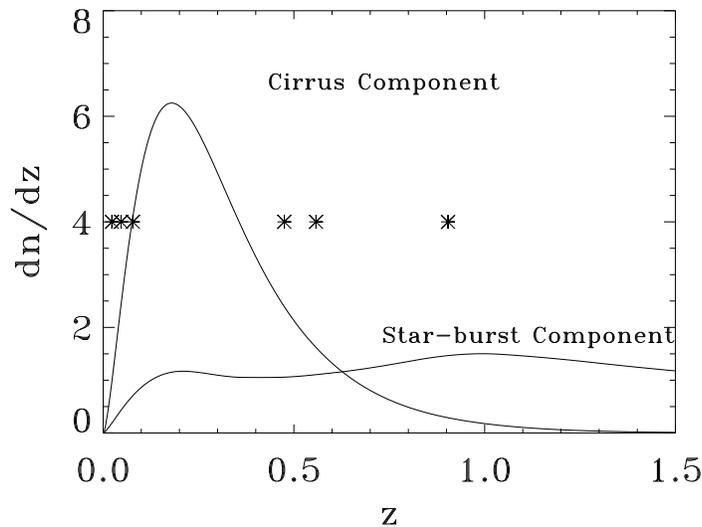, angle=90, width=10cm}
\caption{Expected redshift distribution at 15 $\mu$m within the {\em ISO} HDF areas
(accounting for the area dependent flux limit) for the star-burst
and cirrus component of the PRR model.
Over-plotted at arbitrary $y$ position are the redshifts for
the 6 objects with reliable associations and spectroscopic or
photometric redshifts in Paper IV, the lowest
redshift object has the lowest flux and was excluded from our
count analysis. The probability that these data are drawn from the expected
distribution estimated using the K-S test is 0.41, excluding
the lowest $z$ object. }\label{cppz15}
\end{figure*}

Paper V does not attempt to fit SEDs of the type used for the second
model. However, this model
predicts that the non-AGN 6.7 $\mu$m sources would be 61 per cent
elliptical or S0 and 39 per cent spirals. This is reasonably
consistent with the morphologies and SEDs above.

Only three of the 15 $\mu$m selected sources are associated within the
HDF itself.  Two of these have spiral morphologies and are fitted by
star-burst spectra, while the third has elliptical morphology and a
cirrus spectrum.  With such limited statistics this is reasonably
compatible with both models.  Including the Flanking Field areas we
find in Paper IV
that six of the 15 $\mu$m selected sources have
associations and redshifts (or photometric redshifts).  These are
shown in Figure \ref{cppz15} together with the redshift distribution
from the PRR model. Excluding the lowest redshift object, (which was
excluded from our count analysis as being below the 255mJy flux limit that
was also applied to our model redshift distributions) we find that we
cannot rule out these redshifts being drawn from this expected
distribution (probability 0.41).  This test is hampered by small
number statistics but also assumes that the objects in the flanking
fields with redshifts are not biased in any way.  Clearly obtaining
spectra for the flanking field 15 $\mu$m sources is a high priority.

\section{Implications for Future Surveys and Missions}

The fact that we have reached the {\em ISO} confusion limit at 15 $\mu$ms
with a sensitivity of around 0.2 mJy has important implications for
other surveys and missions, in particular NASA's Small Explorer
Mission, the {\em Wide-Field Infrared Explorer} 
({\em WIRE}, http://www.ipac.caltech.edu/wire/).  {\em WIRE} is due for 
launch in
September 1998 and plans to survey hundreds of square degrees at 12
and 25  $\mu$m.  The {\em WIRE} strategy is divided into three parts: Part I
-- a moderate-depth survey (60 percent of survey time); Part II -- a
deep, confusion-limited survey (30 per cent of the survey time) and
Part III -- an ultra-deep, confusion-distribution measurement.  The
areal coverage and integration times will be designed to achieve these
aims and so the best estimate of the confusion limit is required prior
to planning.  Currently the {\em WIRE} team estimate confusion limits of
between 0.067 mJy and 0.15 mJy at 12 $\mu$m
(http://www.ipac.caltech.edu/WIRE/sensitiv.html) depending on
the evolutionary models.  {\em WIRE} has a 30cm mirror with poorer
resolution than {\em ISO} so our results suggest that the confusion limit
for {\em WIRE} will be brighter than these estimates.

\section{Conclusions}

We have investigated the source counts of our {\em ISO} observations at 6.7
and 15 $\mu$m of the Hubble Deep Field. These are the deepest available
in the mid to far infrared wave-bands, 3 decades fainter than {\em IRAS} at
15 $\mu$m and the first ever presentation of extra-galactic source
counts at 6.7 $\mu$m.  We reach 50$\mu$Jy over 5 arcmin$^2$ at 6.7 $\mu$m, at 15 $\mu$m we cover 5 arcmin$^2$ to a
depth 200$\mu$Jy and around 15 arcmin$^2$ to 400 $\mu$Jy.  In both
bands we cover smaller areas to deeper flux limits.  These data will
be very useful reference points for ongoing {\em ISO} surveys e.g ELAIS (see
e.g. Oliver {\em et al.} 1997), the CAM Deep Survey (see e.g. Elbaz
1997) and the survey of Taniguchi {\em et al.} (1997).  Both the
source counts and a $P(D)$ analysis indicate that we have reached the
confusion limit of {\em ISO} at 15 $\mu$m. This will have important
implications for the construction of mid-infrared surveys with future
space missions such as {\em WIRE}.  At 6.7 $\mu$m we are close to the
confusion limit but have not yet reached it.  At 15 $\mu$m a
no-evolution model is ruled out at $>3\sigma$ providing important
confirmation of the strong evolution seen in {\em IRAS} surveys at a much
greater depth.  The counts appear to be steeper than expected from one
a priori galaxy evolution model at 6.7 $\mu$m although the
significance of this is dependent on our assumptions about the
calibration and noise properties of the data and the FIR SED of the
model which will improve with time.  The $P(D)$ analysis provides
constraints on number count models below the level at which we can
extract reliable source lists.

Further information on the {\em ISO}HDF project can be found on the
{\em ISO}HDF WWW pages: see
http://artemis.ph.ic.ac.uk/hdf/.

\section*{Acknowledgments}
This paper is based on observations with {\em ISO}, an ESA project, with
instruments funded by ESA Member States (especially the PI countries:
France, Germany, the Netherlands and the United Kingdom) and with
participation of ISAS and NASA.
This work was in part supported by PPARC grant no.  GR/K98728
and  EC Network is FMRX-CT96-0068

\section*{ }

\begin{figure*}
\vspace{12cm}
\epsfig{file=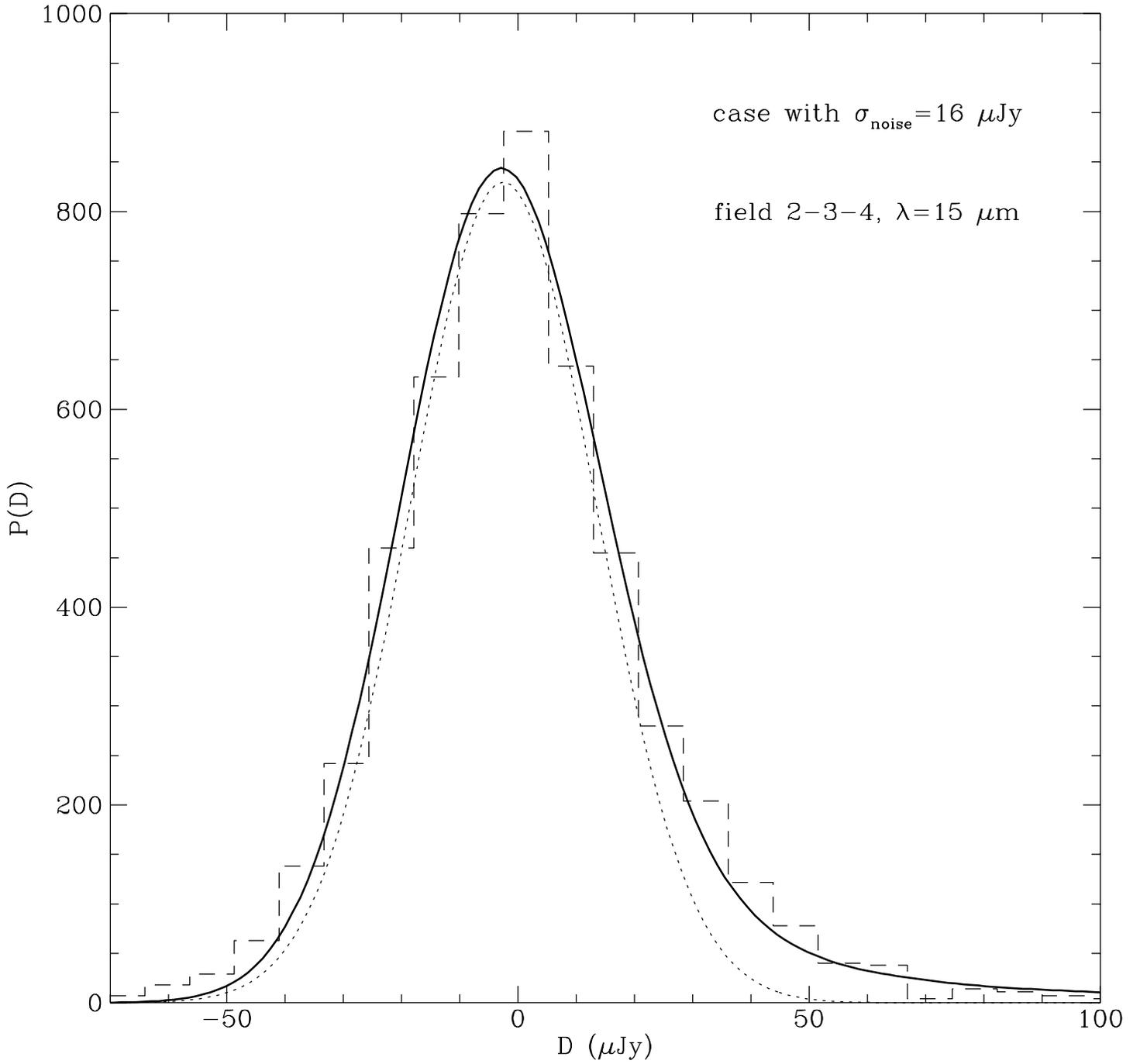, angle=90, width=10cm}

\caption{$P(D)$ analysis at 15 $\mu$m (see text)}
\end{figure*}

\label{lastpage}

\end{document}